\documentclass[12pt,preprint]{aastex}

\begin{document}

\title{The Mid-Infrared Narrow Line Baldwin Effect Revealed by Spitzer}

\author{
Mark~Keremedjiev,\altaffilmark{1}
\email{msk@astro.ufl.edu}
Lei~Hao,\altaffilmark{2,3}
and Vassilis~Charmandaris\altaffilmark{4,5}
}
\altaffiltext{1}{Department of Astronomy, University of Florida, 
Gainesville, FL 32611}
\altaffiltext{2}{Cornell University, Astronomy Department, 
Ithaca, NY 14853-6801}
\altaffiltext{3}{University of Texas, Department of Astronomy, 
Austin, TX 78712}
\altaffiltext{4}{University of Crete, Department of Physics, GR-71003,
 Heraklion, Greece.}
\altaffiltext{5}{IESL/Foundation for Research and Technology - Hellas,
 GR-71110, Heraklion, Greece and Chercheur Associ\'e, Observatoire de
 Paris, F-75014, Paris, France}

\begin{abstract}
  
  We present our discovery of a narrow-line Baldwin effect, an anti-correlation between the equivalent width (EW) of a line and the flux of the associated continuum, in 5-20$\mu$m mid-infrared lines from a sample of 68 Active Galactic Nuclei (AGN), located at z$<$0.5, observed with the Infrared Spectrograph on the {\it Spitzer Space Telescope}. Our analysis reveals a clear anti-correlation between the EW of the [SIV] 10.51$\mu$m, [NeII] 12.81$\mu$m, and [NeIII] 15.56$\mu$m lines and their mid-IR continuum luminosities, while the Baldwin effect for [NeV] 14.32$\mu$m is not as obvious. We suggest that this anti-correlation is driven by the central AGN, not circumnuclear star formation in the host galaxy and present a new method of analyzing this effect in mid-IR lines. We also find that the slope of the narrow-line Baldwin effect in the mid-infrared does not appear to steepen with increasing ionization potential. Examining the dependence of the EW to the Eddington Ratio ($L/L_{Edd}$) we find no strong relationship for mid-IR lines. Our study indicates that the narrow-line mid-infrared Baldwin Effect is quite different from the broad-line optical/UV Baldwin effect and it is possible that the two effects are unrelated. The discovered anti-correlations open new possibilities in the understanding the physics of the ionizing region and the continuum reprocessing by dust.

\end{abstract}

\keywords{galaxies: active --- galaxies: quasars: emission lines --- infrared: galaxies} 

%%%%%%%%%%%%%%%%%%%%%%%%%%%%%%%%%%%%%%%%%%%%%%%%%%%%%%%%%%%%%%%%%%%%%%%%%%

\section{Introduction}

The Baldwin Effect, first discovered by Baldwin (1977), reports the
decrease of the equivalent width (EW) of the broad CIV1549$\AA$
line with increasing UV luminosity in active galactic nuclei.  The
relationship was initially established hoping quasars could be used as
potential standard candles in observational cosmology. Extended
examination of the relationship over the past decades for both quasars
and Seyferts demonstrated that the relationship is not caused by
selection effects, but its cosmological use is limited due to large
scatter (see review of Osmer \& Shields 1999 and Kinney et al. 1990;
Wilkes et al. 1999; Green, Forster \& Kuraszkiewicz 2001; Croom et
al. 2002; Dietrich et al. 2002; Kuraszkiewicz et al. 2002; Shang et
al. 2003). 

Significant correlations also exist between the
continuum emission and the EW in other UV and optical emission
lines including Ly$\alpha$, H$\beta$, CIV, CIII, Ly$\beta$, OIV, OI,
CII, AlIII, CIII, MgII, and SiIV+OIV. It was also found that the slope
of these relationships appear to increase with increasing
ionization potential. In addition, an X-ray Baldwin effect has also
been reported in Fe K$\alpha$ (Iwasawa \& Taniguchi 1993; Nandra et
al. 1997; Page et al. 2004).

The physical origin of this effect is still not clear. A plausible
explanation is the softening of the ionizing continuum shape
with increasing {\it L}, which would lead to weaker emission lines
compared to the local continuum (Baskin \& Laor 2004). However, this
has been challenged by Wilkes et al. (1999), who found no correlation
between any of the UV and optical lines with the X-ray luminosity or
X-ray slope. They suggest a model in which limb darkening and the
projected surface area of an optically thick, geometrically thin disk
combine to cause the Baldwin effect.

Some have also argued that the Eddington ratio $L/L_{Edd}$, a tracer of AGN accretion, may drive the Baldwin effect (Boroson \& Green 1992). The Baldwin effect may then just be a secondary correlation induced by the tendency of more luminous AGN to have a higher $L/L_{Edd}$ (Baskin \& Laor 2004; Shang et al. 2003). Others have argued that the fundamental driver is the mass of the supermassive black hole $M_{BH}$ instead of $L$ or $L/L_{Edd}$ (Warner et al. 2003, 2008).  One further plausible explanation is that metallicity of the gas in the AGN affects the EW of the lines in a way that would generate a Baldwin Effect (Deitrich et al 2002).

Most of the discussion of the Baldwin Effect has focused on broad-lines, but a few papers have also noticed a narrow-line Baldwin effect (Green et al. 2001, Croom et al. 2002, Boroson and Green 1992). Since the narrow-line region in quasars may extend to kpc scales, the physics related to the narrow-line Baldwin effect may be different from those driving the broad-line Baldwin effect (Osmer and Shields 1999), and may simply be due to the covering factor of the narrow-line region (Page et al. 2004). This is manifest in a ``disappearing NLR'' model, where the NLR size is related to the AGN luminosity and highly luminous AGN would have weak or even non-existent NLR (Croom et al. 2002).

Complications arise when moving into the mid-infrared. Several authors
have noted that in many cases, the IR spectra of AGN do not reflect
their optical or UV classifications as reprocessing of the ionizing
radiation by the intervening dust as well as circumnuclear star
formation activity affects the mid-IR spectral features (Lutz et
al. 1998, Laurent et al. 2000, Armus et al. 2007, and Spoon et
al. 2007). As a result, quantifying accurately the AGN contribution to
the infrared or bolometric luminosity of dust enshrouded galaxies is
still largely unanswered problem (see Charmandaris 2008 for a
review). Ascertaining the extent of AGN domination in the mid-IR
presents unique challenges to determining the extent of a Baldwin
Effect. We embarked on the detailed study of the effect using Spitzer
data (Keremedjiev \& Hao 2006) while an analysis using ground based
observations with the VLT/VISIR and comparing to x-ray luminosity has been presented by H\"{o}nig et al. (2008).

In this paper, we report on our discovery of a narrow-line Baldwin
effect in the mid-IR based on Spitzer observations of a large
sample which consists of 68 optically classified AGN. Our observations
and data analysis are presented in Section~2, our results and detected
correlations are shown in Section~3, while the implications of
those are discussed in Section~4.

%%%%%%%%%%%%%%%%%%%%%%%%%%%%%%%%%%%%%%%%%%%%%%%%%%%%%%%%%%%%%%%%%%%%%%%%%%

\section{Observations and Data Reduction}

%%%%%%%%%%%%%%%%%%%%%%%%%%%%%%%%%%%%%%%%%%%%%%%%%%%%%%%%%%%%%%%%%%%%%%%%%%

We compiled a sample of 16 PG Quasars, 33 Seyfert Galaxies, and 19
2MASS AGN with z$<$0.5 observed with the {\it Spitzer Space
  Telescope}'s Infrared Spectrograph (IRS)\footnote{The IRS was a
  collaborative venture between Cornell University and Ball Aerospace
  Corporation funded by NASA through the Jet Propulsion Laboratory and
  the Ames Research Center.}. All objects were observed in the
short-high (SH), short-low (SL) and long-low (LL) IRS modules. The low
resolution modules SL (5.2-14.5$\mu$m) and LL (14.0-38.0$\mu$m) have a
spectral resolution of 64-128, whereas the high resolution module SH
(9.9-19.6$\mu$m) has a resolution of $\sim 600$ (Houck et al. 2004).
For this experiment, our effective coverage is limited by the SH
bandpass. The data were reduced at the Spitzer Science Center (SSC)
using data reduction pipeline 13.2.  The Basic Calibrated Data (BCD)
products from the SSC were used in our analysis.

To obtain the low resolution spectra, we co-added the images of the
same nod position. The background was then subtracted from the
co-added image of the other nod position. The mid-IR spectra were
extracted from these images with the Spectral Modeling,
Analysis, and Reduction Tool (SMART Ver. 5.5.1 Higdon et al. 2004)
using a variable width aperture to recover the diffraction limited
point-spread function.

For high resolution data, the background cannot be subtracted in the same way as in the low resolution cases. Due to the limited size of both apertures of IRS high resolution modules, even point sources tend to fill them making differencing between the two nods impossible. Thus, images of the same nod position were co-added using median averaging and full aperture extractions were conducted in SMART.  The resulting spectra were cosmetically trimmed by eye for order overlap across all orders.  This was done to compensate for the low S/N present in most spectra. Unfortunately separate background observations were not made for our sources, and therefore we cannot apply any background subtraction to our high-resolution data. As a result, the continuum of the high-resolution spectra will always contain a background component.

The spectra were returned to their rest frames using the redshifts of the targets obtained from the NASA Extragalactic Database (NEDS)\footnote{This research has made use  of the NASA/IPAC Extragalactic Database (NED) which is operated by the Jet Propulsion Laboratory, California Institute of Technology, under contract with the National Aeronautics and Space Administration.}. The continuum flux was calculated from the local continuum in the low resolution data.  Calculation of the continuum luminosity was done with distances determined from z where we assumed a value of 73 $km/s/Mpc$ was used for $H_{0}$ (Riess et al 2005).

Line strengths were measured from the high-resolution spectra. Since a significant number of our sources only have SH data ($9.9\mu m<\lambda_{SH}<19.6\mu m$), we will focus on studying the narrow-line Baldwin Effect in strong emission lines in the SH wavelength range. More specifically these lines and their corresponding ionization potential are: [SIV] 10.501$\mu$m (47.22eV), [NeII] 12.814$\mu$m (40.96eV), [NeV] 14.320$\mu$m (126.25eV), and [NeIII] 15.555$\mu$m (63.42eV). Line strengths were determined by fitting a Gaussian function to the feature at the central wavelength. The local continuum was subtracted off with a linear fit. 

Equivalent widths were measured by dividing the line strength measured in high-resolution by the continuum strength measured in low-resolution.  This was done because the low-resolution data do not contain the background flux present in the high resolution data and thus they more accurately reflect the actual continuum flux. In Table 1, we present line strengths, continuum values, and EW for all our objects as well as redshifts and luminosity distances.

%+++++++++++++++++++++++++++++++++++++++++++++++++++++++++++++++++++++
\begin{deluxetable}{lccccccccccccccc} % Table 1
\tabletypesize{\tiny}
\tablecaption{{Various measured properties for the galaxies in our sample. $D_L$ is given in Mpc, fluxes are in $(10^{-21}  W cm^{-2})$, EW are measured in $(\mu m)$, and continuum values are measured in $(10^{-21}  W cm^{-2} \mu m^{-1})$.}\label{tab1}}
\tablewidth{0pt}
\rotate
\tablehead{\colhead{} & \colhead{} & \colhead{} & \colhead{} & \colhead{[SIV]} & \colhead{} & \colhead{} & \colhead{[NeII]} & \colhead{} & \colhead{} & \colhead{[NeV]} & \colhead{} & \colhead{} & \colhead{[NeIII]} & \colhead{} & \colhead{$5.5\mu m$} \\ 
\cline{4-6} \cline{10-12} \cline{16-16}\\
\colhead{Name} & \colhead{z} & \colhead{$D_L$} & \colhead{Flux} & \colhead{EW} & \colhead{Cont.} & \colhead{Flux} & \colhead{EW} & \colhead{Cont.} & \colhead{Flux} & \colhead{EW} & \colhead{Cont.} & \colhead{Flux} & \colhead{EW} & \colhead{Cont.} & \colhead{Cont.} %\\

%\colhead{} & \colhead{} & \colhead{(Mpc)} & \colhead{$10^{-21}  W/cm^2$} & \colhead{$\mu m$} & \colhead{$10^{-21}  W cm^{-2} \mu m^{-1}$} & \colhead{$10^{-21}  W/cm^2$} & \colhead{$\mu m$} & \colhead{$10^{-21}  W cm^{-2} \mu m^{-1}$} & \colhead{$10^{-21}  W/cm^2$} & \colhead{$\mu m$} & \colhead{$10^{-21}  W cm^{-2} \mu m^{-1}$} & \colhead{$10^{-21}  W/cm^2$} &\colhead{$\mu m$} &  \colhead{$10^{-21}  W cm^{-2} \mu m^{-1}$} & \colhead{$10^{-21}  W cm^{-2} \mu m^{-1}$}
} 
\startdata
2MASSJ000703 & 0.114 & 482 & 2.32 & 0.0361 & 64.1 & 2.29 & 0.0377 & 60.7 & 2.69 & 0.0503 & 53.4 & 5.29 & 0.105 & 50.5 & 115 \\
2MASSJ005055 & 0.136 & 575 & 2.32 & 0.0241 & 96.3 & $<$1.11 & $<$0.0122 & 90.8 & 4.27 & 0.0500 & 85.4 & N/A & N/A & N/A & 170 \\
2MASSJ010835 & 0.285 & 1200 & $<$0.6910 & $<$0.0285 & 24.2 & 1.66 & 0.0643 & 25.8 & $<$0.616 & $<$0.0264 & 23.3 & $<$1.73 &  $<$0.0872 & 19.8 & 86.5 \\
2MASSJ015721 & 0.213 & 900 & $<$0.6450 & $<$0.00778 & 83.0 & 1.57 & 0.0182 & 86.3 & $<$0.739 & $<$0.00946 & 78.1 & $<$0.853 &  $<$0.0119 & 71.5 & 163 \\
2MASSJ034857 & 0.140 & 592 & $<$1.109 & $<$0.00690 & 161 & 0.952 & 0.00595 & 160 & $<$0.912 & $<$0.00667 & 137 & $<$0.977 &  $<$0.00825 & 118 & 570 \\
2MASSJ091848 & 0.210 & 887 & 2.26 & 0.0264 & 85.7 & 0.982 & 0.0122 & 80.7 & 4.48 & 0.0626 & 71.7 & 2.60 & 0.0396 & 65.8 & 198 \\
2MASSJ102724 & 0.149 & 630 & $<$0.9020 & $<$0.0131 & 68.7 & $<$0.740 & $<$0.0113 & 65.8 & $<$0.971 & $<$0.0165 & 58.8 & $<$1.22 &  $<$0.0218 & 55.8 & 161 \\
2MASSJ105144 & 0.231 & 976 & 1.66 & 0.0671 & 24.7 & 1.76 & 0.0693 & 25.5 & 1.61 & 0.0643 & 25.0 & 3.02 & 0.122 & 24.7 & 55.3 \\
2MASSJ130005 & 0.080 & 338 & $<$0.878 & $<$0.00357 & 246 & 1.67 & 0.00870 & 192 & $<$0.706 & $<$0.00432 & 163 & 1.45 & 0.00978 & 148 & 526 \\
2MASSJ140251 & 0.187 & 790 & $<$0.644 & $<$0.0160 & 40.3 & $<$0.533 & $<$0.01867 & 28.7 & $<$0.811 & $<$0.0352 & 23.1 & $<$0.791 &  $<$0.0376 & 21.0 & 91.0 \\
2MASSJ145331 & 0.139 & 587 & $<$0.600 & $<$0.0158 & 38.1 & 4.77 & 0.0474 & 101 & $<$0.774 & $<$0.00893 & 86.7 & 4.75 & 0.0649 & 73.2 & 295 \\
2MASSJ150113 & 0.258 & 1.09e3 & $<$0.617 & $<$0.0148 & 41.9 & $<$0.746 & $<$0.0187 & 39.9 & $<$0.797 & $<$0.0219 & 36.4 & N/A  &   N/A    & 0 & 90.3 \\
2MASSJ151653 & 0.190 & 803 & $<$0.692 & $<$0.00277 & 250 & $<$0.722 & 0.00468 & 154 & $<$0.717 & $<$0.00587 & 122 & 2.19 & 0.0202 & 109 & 534 \\
2MASSJ163700 & 0.211 & 892 & $<$0.780 & $<$0.0311 & 25.0 & $<$0.743 & $<$0.0279 & 26.7 & $<$0.810 & $<$0.0411 & 19.7 & $<$1.19 &  $<$0.0611 & 19.4 & 50.1 \\
2MASSJ165939 & 0.170 & 718 & 2.01 & 0.0143 & 141 & 1.44 & 0.0119 & 121 & 2.43 & 0.0229 & 106 & 4.62 & 0.0463 & 99.7 & 210 \\
2MASSJ171442 & 0.163 & 689 & $<$0.623 & $<$0.0165 & 37.7 & 1.32 & 0.0468 & 28.3 & $<$0.734 & $<$0.0335 & 21.9 & $<$0.605 &  $<$0.0285 & 21.2 & 79.6 \\
2MASSJ222221 & 0.211 & 892 & $<$0.706 & $<$0.0116 & 60.9 & $<$0.578 & $<$0.0129 & 45.0 & 0.903 & 0.0241 & 37.5 & $<$1.05 &  $<$0.0290 & 36.2 & 113 \\
2MASSJ222554 & 0.147 & 621 & $<$0.605 & $<$0.0157 & 38.5 & 0.573 & 0.0128 & 44.6 & 1.02 & 0.0266 & 38.2 & 1.02 & 0.0300 & 33.8 & 98.9 \\
2MASSJ234449 & 0.199 & 841 & 1.25 & 0.0217 & 57.8 & $<$0.772 & $<$0.0171 & 45.2 & 1.90 & 0.0477 & 39.9 & 2.12 & 0.0565 & 37.5 & 109 \\
3C120 & 0.033 & 140 & 24.7 & 0.0410 & 603 & 8.20 & 0.0175 & 468 & 18.3 & 0.0411 & 444 & 29.9 & 0.0663 & 451 & 990 \\
3C273 & 0.158 & 669 & 3.55 & 0.00450 & 788 & 1.56 & 0.00279 & 558 & 2.85 & 0.00576 & 495 & 5.78 & 0.0121 & 480 & 1.93e3 \\
3C445 & 0.0562 & 238 & 2.39 & 0.00573 & 417 & 2.90 & 0.00988 & 293 & 3.65 & 0.0149 & 244 & 3.82 & 0.0153 & 250 & 791 \\
ESO103-g035 & 0.0133 & 56.2 & 7.35 & 0.00768 & 957 & 20.4 & 0.0120 & 1.69e3 & 11.1 & 0.00633 & 1.75e3 & 23.9 & 0.0145 & 1.65e3 & 1.63e3 \\
Fairall9 & 0.0470 & 199 & $<$1.52 & $<$0.00226 & 674 & 3.26 & 0.00648 & 503 & 1.71 & 0.00382 & 448.6 & 4.51 & 0.0106 & 426 & 1.07e3 \\
H1846-786 & 0.0743 & 314 & $<$1.05 & $<$0.0109 & 96.2 & 1.45 & 0.0163 & 89.3 & $<$0.716 & $<$0.00770 & 93.0 & 1.63 & 0.0182 & 89.8 & 170 \\
IC4329a & 0.0161 & 67.8 & 22.4 & 0.0101 & 2.21e3 & 21.9 & 0.0110 & 1.99e3 & 25.6 & 0.0133 & 1.93e3 & 42.5 & 0.0233 & 1.82e3 & 3.02e3 \\
IC5135 & 0.0162 & 68.2 & 4.62 & 0.0118 & 393 & 65.3 & 0.0989 & 660 & 9.01 & 0.0156 & 578 & 20.0 & 0.0329 & 606 & 502 \\
IRAS 07145-2914 & 0.00566 & 23.9 & 94.4 & 0.198 & 478 & 50.8 & 0.0673 & 756 & 62.7 & 0.0727 & 863 & 142 & 0.156 & 911 & 648 \\
I Zw 1 & 0.0611 & 258 & $<$2.34 & $<$0.00167 & 1.40e3 & 2.93 & 0.00293 & 1.00e3 & 2.57 & 0.00297 & 863 & 5.22 & 0.00630 & 829 & 1.88e3 \\
M-6-30-15 & 0.00775 & 32.8 & 7.30 & 0.0100 & 730 & 3.67 & 0.00572 & 641 & 5.08 & 0.00810 & 628 & 6.76 & 0.0117 & 578 & 1.25e3 \\
Mrk1066 & 0.0120 & 50.8 & 10.5 & 0.0240 & 439 & 110 & 0.141 & 776 & 10.0 & 0.0142 & 704 & 46.7 & 0.0616 & 759 & 607 \\
Mrk3 & 0.0135 & 57.1 & 60.6 & 0.0504 & 1.20e3 & 100 & 0.0612 & 1.64e3 & 64.6 & 0.0357 & 1.81e3 & 182 & 0.0930 & 1.96e3 & 1.10e3 \\
Mrk509 & 0.0344 & 145 & 9.24 & 0.0142 & 649 & 12.0 & 0.0225 & 535 & 6.23 & 0.0127 & 489 & 16.8 & 0.0332 & 506 & 1.23e3 \\
NGC1275 & 0.0176 & 74.2 & 1.38 & 7.00E-4 & 1.96e3 & 47.6 & 0.0270 & 1.76e3 & 3.07 & 0.00174 & 1.77e3 & 20.3 & 0.0108 & 1.88e3 & 1.08e3 \\
NGC1386 & 0.0029 & 12.3 & 25.9 & 0.0443 & 585 & 15.4 & 0.0166 & 925 & 33.8 & 0.0376 & 900 & 35.4 & 0.0436 & 812 & 1.44e3 \\
NGC2110 & 0.00779 & 32.9 & 6.64 & 0.0115 & 580 & 56.9 & 0.116 & 489 & 5.21 & 0.0141 & 371 & 45.4 & 0.107 & 424 & 845 \\
NGC2273 & 0.00614 & 25.9 & 3.73 & 0.00660 & 565 & 42.8 & 0.0573 & 747 & 4.74 & 0.00701 & 676 & 19.5 & 0.0291 & 670 & 970 \\
NGC3081 & 0.00798 & 33.7 & 33.3 & 0.0751 & 444 & 12.5 & 0.0239 & 525 & 31.5 & 0.0576 & 547 & 35.4 & 0.0606 & 584 & 445 \\
NGC4151 & 0.00332 & 14.0 & 92.7 & 0.0202 & 4.58e3 & 134 & 0.0328 & 4.10e3 & 77.1 & 0.0195 & 3.94e3 & 204 & 0.0510 & 4.00e3 & 7.07e3 \\
NGC4388 & 0.00842 & 35.6 & 53.9 & 0.112 & 480 & 81.2 & 0.0913 & 889 & 46.0 & 0.0491 & 938 & 98.0 & 0.103 & 953 & 1.04e3 \\
NGC4507 & 0.00842 & 35.6 & 9.40 & 0.00796 & 1.18e3 & 33.5 & 0.0315 & 1.06e3 & 10.6 & 0.00992 & 1.07e3 & 32.1 & 0.0301 & 1.07e3 & 2.33e3 \\
NGC4939 & 0.0104 & 43.9 & 18.7 & 0.212 & 88.2 & 7.74 & 0.0693 & 112 & 12.2 & 0.104 & 117 & 25.3 & 0.188 & 134 & 105 \\
NGC4941 & 0.0037 & 15.6 & 8.77 & 0.0658 & 133 & 14.7 & 0.0976 & 150 & 8.21 & 0.0509 & 161 & 22.4 & 0.124 & 181 & 157 \\
NGC5135 & 0.0137 & 58.0 & 12.3 & 0.0215 & 573 & 93.0 & 0.0986 & 942 & 12.3 & 0.0174 & 708 & 41.3 & 0.0588 & 702 & 619 \\
NGC5347 & 0.00779 & 32.9 & $<$1.49 & $<$0.00268 & 556 & 5.19 & 0.00812 & 638 & 2.01 & 0.00322 & 625 & 4.43 & 0.00733 & 605 & 486 \\
NGC5506 & 0.00618 & 26.1 & 37.8 & 0.0219 & 1.73e3 & 48.4 & 0.0184 & 2.64e3 & 28.1 & 0.0113 & 2.50e3 & 69.7 & 0.0316 & 2.20e3 & 5.75e3 \\
NGC5548 & 0.0172 & 72.6 & 4.42 & 0.00849 & 520 & 9.13 & 0.0202 & 453 & 4.13 & 0.00965 & 428 & 9.89 & 0.0232 & 426 & 558 \\
NGC5643 & 0.00400 & 16.9 & 23.2 & 0.0410 & 566 & 40.3 & 0.0455 & 887 & 25.4 & 0.0297 & 855 & 55.0 & 0.0615 & 895 & 561 \\
NGC7172 & 0.00868 & 36.7 & 4.99 & 0.0363 & 138 & 33.0 & 0.0762 & 433 & 8.63 & 0.0270 & 320 & 16.5 & 0.0671 & 246 & 1.05e3 \\
NGC7213 & 0.00595 & 25.1 & $<$1.26 & $<$0.00182 & 690 & 25.4 & 0.0578 & 439 & $<$0.878 & $<$0.00225 & 390 & 12.9 & 0.0311 & 416 & 871 \\
NGC7314 & 0.00476 & 20.1 & 17.3 & 0.120 & 144 & 9.89 & 0.0483 & 205 & 16.4 & 0.0773 & 212 & 20.9 & 0.100 & 209 & 230 \\
NGC7469 & 0.0163 & 69.0 & 9.65 & 0.00465 & 2.07e3 & 207 & 0.0883 & 2.34e3 & 15.0 & 0.00740 & 2.02e3 & 34.8 & 0.0162 & 2.15e3 & 2.08e3 \\
PG0804+761 & 0.100 & 423 & 2.16 & 0.00529 & 408 & $<$0.521 & $<$0.00222 & 235 & $<$0.456 & $<$0.00250 & 182 & 1.31 & 0.00739 & 177 & 840 \\
PG1119+120 & 0.0502 & 212 & 2.21 & 0.00948 & 233 & 0.613 & 0.00315 & 195 & 2.19 & 0.0126 & 174 & 2.80 & 0.0164 & 171 & 296 \\
PG1211+143 & 0.0809 & 342 & $<$0.676 & $<$0.00139 & 487 & $<$0.445 & $<$0.00134 & 332 & 1.39 & 0.00478 & 291 & 1.13 & 0.00394 & 285 & 849 \\
PG1351+640 & 0.0882 & 373 & 2.49 & 0.00494 & 503 & 2.59 & 0.00869 & 298 & $<$0.607 & $<$0.00222 & 274 & 3.28 & 0.0110 & 300 & 427 \\
PG1501+106 & 0.0364 & 154 & 9.02 & 0.0200 & 451 & 4.43 & 0.0104 & 427 & 9.00 & 0.0222 & 406 & 12.6 & 0.0311 & 406 & 596 \\
PG2130+099 & 0.0630 & 266 & 4.37 & 0.0108 & 404 & 1.93 & 0.00576 & 336 & 3.80 & 0.0131 & 291 & 6.19 & 0.0226 & 274 & 787 \\
PG0838+770 & 0.131 & 554 & $<$0.260 & $<$0.00319 & 81.6 & $<$0.458 & $<$0.00701 & 65.4 & $<$0.401 & $<$0.00664 & 60.3 & $<$0.984 & $<$0.0160 & 61.4 & 116 \\
PG1302-102 & 0.278 & 1180 & 2.15 & 0.0151 & 142 & $<$0.349 & $<$0.00295 & 118 & 0.480 & 0.00439 & 109 & $<$0.851 & $<$0.00347 & 245 & 174 \\
PG1411+442 & 0.0896 & 379 & 1.10 & 0.00421 & 262 & $<$0.306 & $<$0.001780 & 172 & 0.964 & 0.00676 & 143 & 1.10 & 0.00851 & 129 & 530 \\
PG1426+015 & 0.0865 & 365 & 1.75 & 0.00630 & 277 & 1.23 & 0.00615 & 200 & 1.44 & 0.00809 & 178 & 2.63 & 0.0154 & 171 & 505 \\
PG1440+356 & 0.0791 & 334 & 1.73 & 0.00848 & 204 & 4.52 & 0.0286 & 158 & 1.83 & 0.0137 & 133 & 3.74 & 0.0279 & 134 & 685 \\
PG1613+658 & 0.129 & 545 & 1.31 & 0.00525 & 250 & 4.39 & 0.0236 & 186 & 2.33 & 0.0143 & 162 & 3.83 & 0.0240 & 159 & 488 \\
PG1626+554 & 0.133 & 562 & $<$0.179 & $<$0.00330 & 54.4 & $<$0.130 & $<$0.00477 & 27.2 & 0.401 & 0.0183 & 22.0 & $<$0.157 & $<$0.00772 & 20.4 & 115 \\
PG2214+139 & 0.0658 & 278 & 0.724 & 0.00321 & 226 & 1.26 & 0.00894 & 141 & $<$0.424 & $<$0.00391 & 109 & 0.781 & 0.00756 & 103 & 522 \\
PG2251+113 & 0.326 & 1380 & 0.854 & 0.0126 & 67.8 & $<$0.268 & $<$0.00515 & 52.0 & 0.868 & 0.00710 & 122 & 5.58 & 0.0254 & 220 & 161 \\
PG2349-014 & 0.174 & 735 & 1.88 & 0.0197 & 95.7 & 1.76 & 0.0215 & 81.9 & 0.938 & 0.0114 & 82.3 & 2.53 & 0.0305 & 83.1 & 205 \\
\enddata
\end{deluxetable}
%+++++++++++++++++++++++++++++++++++++++++++++++++++++++++++++++++++++

%%%%%%%%%%%%%%%%%%%%%%%%%%%%%%%%%%%%%%%%%%%%%%%%%%%%%%%%%%%%%%%%%%%%%

\section{Results}

%+++++++++++++++++++++++++++++++++++++++++++++++++++++++++++++++++++++
\begin{figure*} % Fig. 1
\begin{center}
\resizebox{\hsize}{!}{\includegraphics[angle=90]{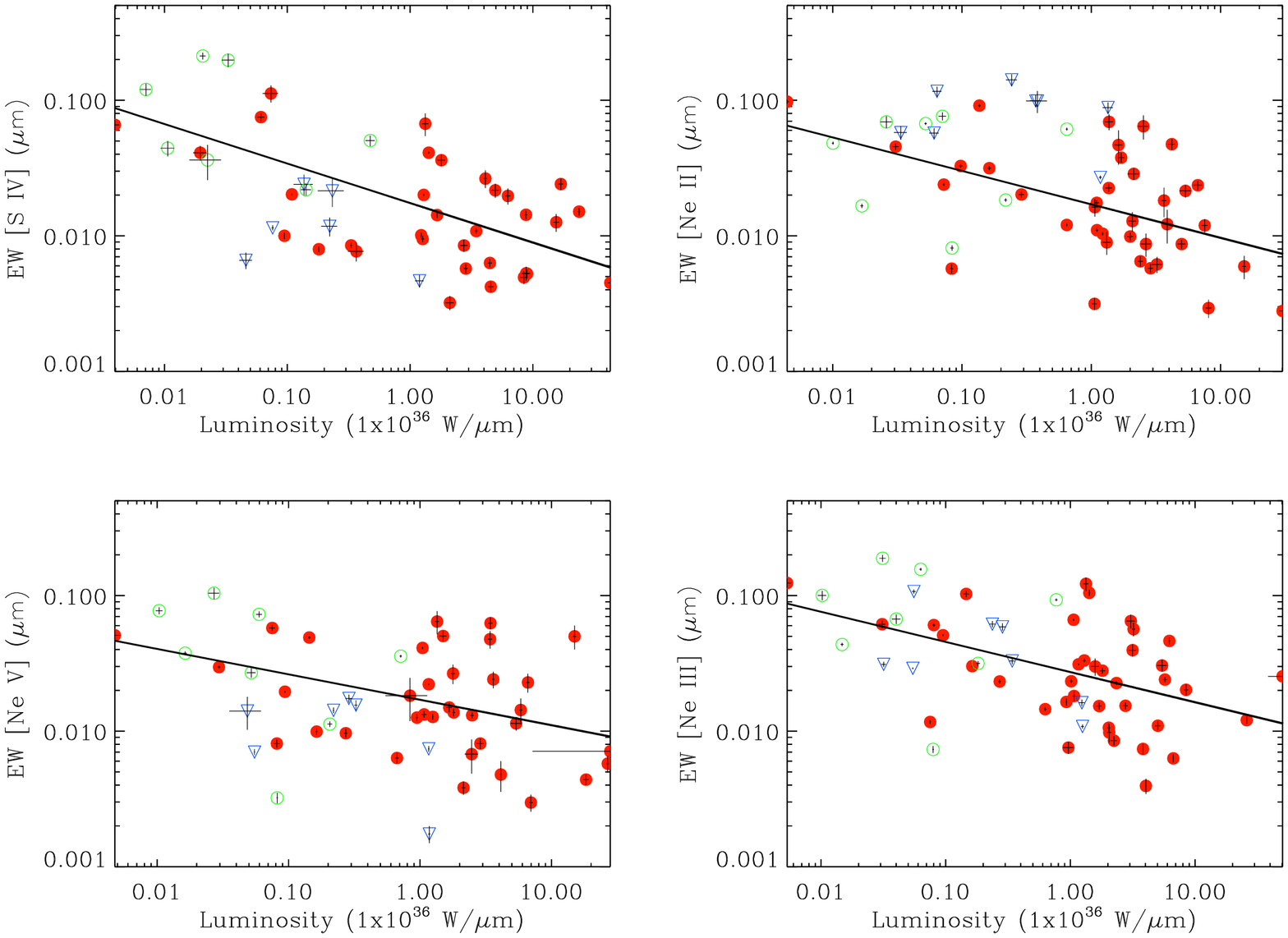}}

\caption{Plots of the EW of [SIV], [NeII], [NeV], and [NeIII] as a function of the continuum monochromatic luminosity next to the line. The green open circles denote Seyfert 2 galaxies while the red filled circles indicate the Seyfert 1 galaxies. Starburst galaxies are marked in blue open triangles. The data are overplotted with a lines representing the least-squares fits.
\label{fig1}}
\end{center}
\end{figure*}
%+++++++++++++++++++++++++++++++++++++++++++++++++++++++++++++++++++++

In Figure 1, we present the EW of the four lines as a function of the nearby mid-IR monochromatic continuum luminosity. A narrow-line Baldwin Effect, where EW declines as luminosity increases, appears to manifest itself in all four lines across six orders of magnitude in luminosity. As mentioned earlier, our sample contains both type I (quasars and Seyfert 1s) and type II (mostly Seyfert 2s) AGN and we include both types in our analysis. In all figures we indicate type 1 objects with filled red circles and type 2 with open green circles. We do not see any evidence that either type 1 or type 2 AGN deviate from the trend and while type 2 AGN are generally in the lower mid-IR luminosity range, they do not skew the overall Baldwin effect. The blue open triangles denote galaxies with significant circumnuclear starburst contribution, classified in a manner described below. We exclude them from further correlation analysis.

In Fig. 1 we also include the error bars to our data. The
uncertainties in the continuum luminosity were generated by
calculating the variance in the local continua. No assumptions were
made about uncertainties in the redshift measurements and therefore
our errors are likely lower limits. The flux density uncertainty was
a combination of variance in the local high resolution continuum as
well as error in the profile fitting. Only lines detected with confidence
greater than $3\sigma_{flux}$, were included in our analysis. The
error in EW was the propagation of the noise both in low resolution
$\sigma_{continuum}$, and in high resolution $\sigma_{flux}$. For
nearly all our data, the uncertainty was $<$10\% indicating good
quality of our measurements. The reason why the error bars appear so
small on the plot has much to do with the large dynamic range in luminosities
and EW measured in our sample.

We used the conventional definition of the Baldwin Effect expressed as

\begin{equation}
W_{\lambda}=\alpha L_{\lambda}^{\beta}
\end{equation}

where $W_{\lambda}$ indicates the EW and $\beta$ the slope of the anti-correlation. We performed a least squares fit to the data in logarithmic space and to test the linear correlation of the log of EW versus log of luminosity, a Spearman rank correlation was employed.  The best fits to the data are also plotted in Figure 1. 

In three lines, [SIV], [NeII], and [NeIII] we find a greater than $3\sigma$ significance in the anti-correlation. For [SIV], the Spearman rank correlation strength is -0.60 with a null-hypothesis value of $4.16\times10^{-5}$. The scatter is comparable to the CIV Baldwin effect (Kinney et al. 1990) and the least-squares slope is $-0.29\pm0.05$; steeper than the CIV Baldwin effect which is near -0.17 (Kinney et al 1990, Wilkes et al 1999, Osmer and Shields 1999, and Laor et al 1995). The [NeII] anti-correlation also manifests itself with a Spearman rank value of -0.48, significance of $8.88\times10^{-4}$, and a slope of $-0.25\pm0.06$. For [NeIII], the correlation is -0.46, $1.17\times10^{-3}$ significance, and a slope of $-0.22\pm0.06$. 

The significance of these correlations were tested in two different ways. The first was by calculating the number of deviations from the null-hypothesis value for the Spearman rank coefficient. The second was by running Monte Carlo simulations where we randomly assigned our EW values to our continuum measurements. This was done 100,000 times for each line and the resulting correlations were noted. In all cases where we claim to have a greater than $>3\sigma$ significance, the Monte Carlo simulations provided a second, independent verification.

Unlike the first three lines, when we examine the [NeV] line we find that the correlation strength is just -0.41 ($5.34\times10^{-3}$ significance) with a slope of $-0.19\pm0.06$. The $2.5\sigma$ significance and low correlation value are considered marginal. As we will discuss in the following sections, this was rather unexpected since [NeV], due to its high excitation potential, is telltale sign of AGN activity. Since the Baldwin Effect is, at least in the optical/UV, directly coupled with the AGN activity one would expect that it would be more prominent in strong mid-IR AGN lines as well.

%+++++++++++++++++++++++++++++++++++++++++++++++++++++++++++++++++++++

\subsection{Eddington Luminosities}

In order to examine the possibility of the Eddington ratio ($L/L_{Edd}$) as a driver for the mid-IR narrow line Baldwin effect, we searched the literature and calculated $L/L_{Edd}$ for 29 of our sources that had measured bolometric luminosities and black hole masses.% The $L/L_{Edd}$ values along with references are presented in Table 1.

Our analysis revealed no strong relationship between Eddington Luminosity and EWs for mid-IR lines. The Spearman rank correlation values we find are -0.21 for [SIV], -0.45 for [NeII], -0.36 for [NeV], and -0.36 for [NeIII]. Their corresponding null-hypothesis values are 0.320, 0.032, 0.070, and 0.063. The low correlation values coupled with the fact that none of relations have $>3\sigma$ significance indicates that the Eddington ratio is not a driving factor for the weakening of lines in the narrow line region.

%+++++++++++++++++++++++++++++++++++++++++++++++++++++++++++++++++++++

\subsection{Slopes of the Baldwin Effect}

Previous studies of the broad line Baldwin effect have revealed that as the ionization potential increases, the steepness of the anti-correlation also increases (Osmer \& Shields 1999, Wu et al. 1983; Kinney et al. 1987; Kinney, Rivolo, \& Koratkar 1990; Badwin et al. 1989; Zheng et al. 1997; Dietrich et al 2002). This was first noticed in the comparisons between low ionization lines such as Mg II and $Ly\alpha$ and high ionization lines such as C IV and O VI.

To examine the relationship between slope and ionization potential in our study, we have compiled a list of $\beta$ slopes from the literature (Kinney, Rivolo, \& Koratkar 1990; Zheng, Kriss, \& Davidsen 1995; Dietrich et al. 2002, Croom et al. 2002) and also included data from our sample. The result is given in Fig. 2 and demonstrates that the values from our narrow-line study of AGN in the mid-IR do not appear confirm earlier assessment that the broad-line slope of the Baldwin effect becomes steeper when one considers lines of increasing ionization potential. Further confirmation of this finding is given by H\"{o}nig et al. (2008), who presented anti-correlations between mid-IR EW's and 2-10 keV luminosity. They found that the steepness of their anti-correlations were also independent of ionization potentials. 

Comparing to narrow-line optical data, the results are inconclusive.  Croom et al. (2002) only reported strong anti-correlations for two lines, [NeV] and [OII] (plotted as blue, open diamonds in Fig. 2), but felt that [OII] might be contaminated by host galaxy emission. Therefore, there are not enough data points to draw definitive conclusions. They suspected that optical, narrow line slopes steepen with ionization potential and, if so, then the physical processes driving mid-IR narrow lines may be quite different from their optical counterparts.

%+++++++++++++++++++++++++++++++++++++++++++++++++++++++++++++++++++++
\begin{figure} % Fig. 2
\begin{center}
\resizebox{\hsize}{!}{\includegraphics[angle=90]{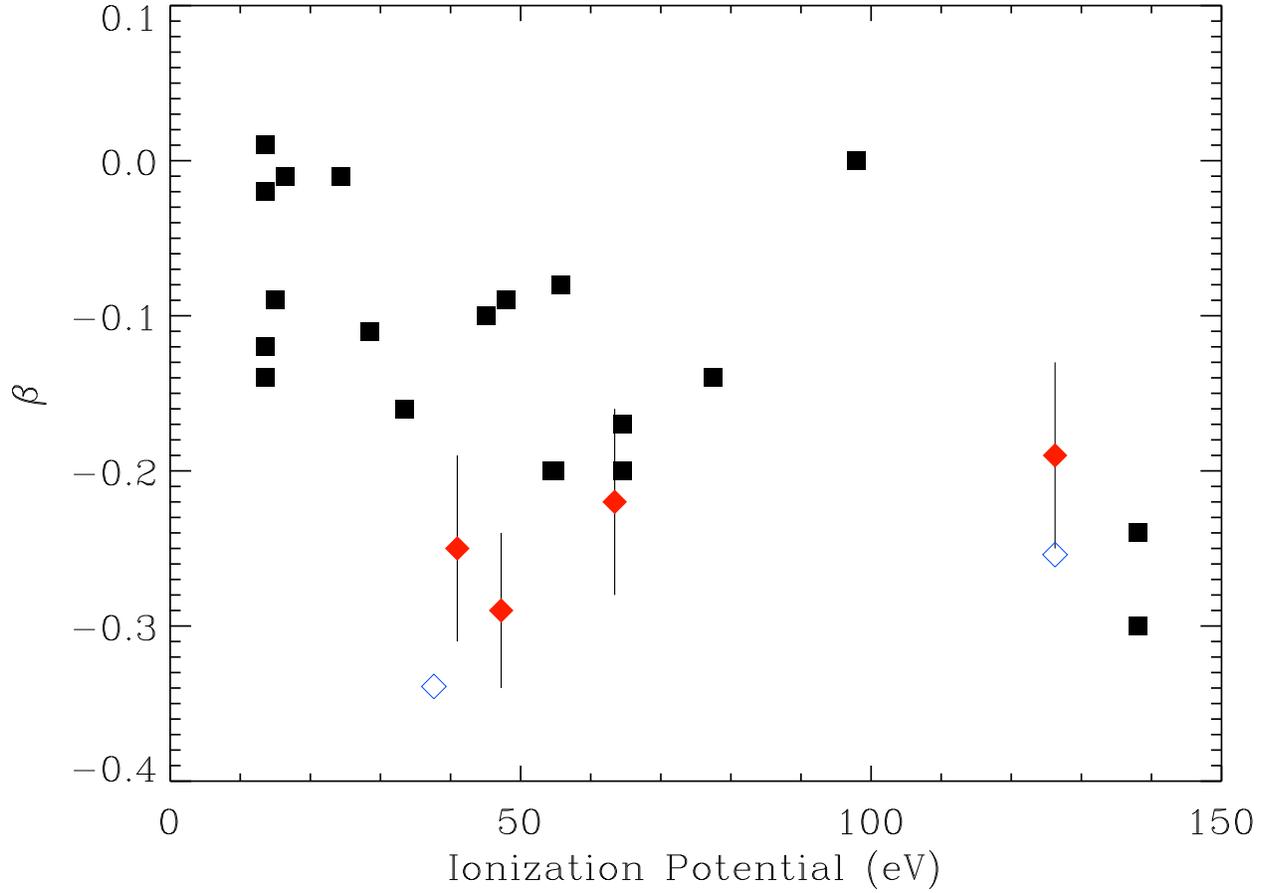}}
\caption{
Plot of ionization potential versus anti-correlation slope, $\beta$. The red filled diamonds denote narrow-line data from this paper and have their associated error bars. The blue open diamonds are narrow line data from Croom et al. (2002), and black squares are points taken from various sources and represent broad-line data.
\label{fig2}}
\end{center}
\end{figure}
%+++++++++++++++++++++++++++++++++++++++++++++++++++++++++++++++++++++

%%%%%%%%%%%%%%%%%%%%%%%%%%%%%%%%%%%%%%%%%%%%%%%%%%%%%%%%%%%%%%%%%%%%%

\section{Discussion}

In the original Baldwin effect, the EW of the lines was anti-correlated with the strength of the UV continuum emission, and the latter was used as tracer of the AGN intensity. This technique is not as direct in the mid-IR because the local continuum may not reflect AGN intensity and could introduce some scatter in the correlations we present in Fig. 1. It is well known that in the infrared, the spectral signatures of an AGN can be severely blended by emission originating from circumnuclear starbursts (see Laurent et al. 2000, Armus et al. 2007 and references therein). This is only partially due to limited spatial resolution of infrared telescopes and focal plane arrays.  More importantly it is the intervening dust which fully reprocesses the intrinsic radiation from the sources - both massive stars and/or an accretion disk - and re-emits it in the infrared. This may lead to a difference between the optical and mid-IR classification of a source.  To address this issue a number of diagnostic methods have been developed of the years in order to obtain a robust AGN/starburst classification of mid-IR spectra (see Lutz et al. 1998, Laurent et al. 2000, Armus et al. 2007, Spoon et al. 2007, Charmandaris 2008, Nardini et al. 2008).

It is important for our analysis to explore whether the observed anti-correlation we find is indeed related to the strength of the AGN. Ideally one could address this by examining the correlation between the line EW and the strength of the X-ray emission (see Hoenig et al. 2008). However, these data are not available for our sample so instead, one can examine how strongly the AGN contributes to the mid-IR continuum emission near the lines versus how much is contaminated by circumnuclear star formation.

It is widely believed that [NeV] 14.32 $\mu$m, due to its high ionization potential, originates in the NLR (see Gorjian et al. 2007 and references there in). So one would have to examine whether most of the emission from [NeII], [NeIII], and [SIV] in AGN also come from the same region. Gorjian et al (2007) found that both the [NeV] 14.32 $\mu$m and [NeIII] 15.6$\mu$m are strongly correlated and deduced that the two must be produced in the same region. Testing the correlation between [NeV] and [NeIII] in our sample confirms this results at $>5\sigma$ confidence level. We also examined whether [SIV] and [NeII] correlate with [NeV] and found that as with [NeIII], both possessed $>5\sigma$ relationship with very little scatter. Altogether, this suggests that the four lines investigated in this paper are likely to arise from the same region.

To estimate the contribution of massive star formation in the
circumnuclear regions of the AGN sampled by the IRS slits, we used
two diagnostics which have been proposed by Genzel et al. (1998) and
have also been applied by Armus et al. (2007) in a study of local
ultralumninous infrared galaxies. The first diagnostic was to use
the [NeV]/[NeII] ratio. This ratio is useful because it is very
difficult for stellar sources to produce a significant number of
photons capable of ionizing [NeV]. As a result a high [NeV]/[NeII] 
line ratio indicates the presence an AGN which is dominant in the mid-IR 
(Lutz et al. 1998).  

A second diagnostic is to use the EW of the 6.2$\mu m$ emission feature. When this feature is strong, it indicates the presence of Polycyclic Aromatic Hysdrocarbons (PAHs) which are mainly produced in photo-dissociation regions (PDRs) when they are excited by the adjacent star forming regions.  Therefore, in Figure 3 we plot the EW$(6.2\mu m)$ versus [NeV]/[NeII] similar to Figure 5 in Armus et al. (2007). From Figure 3 we see that there are eight galaxies (NGC 7469, IC 5135, NGC 5135, Mrk1066, NGC 1275, NGC 2110, NGC 2273, and NGC 7213) that have both a low [NeV]/[NeII] value and strong PAH emission. As mentioned before, we flagged them as starburst galaxies and did not use them in our correlation analysis.

%+++++++++++++++++++++++++++++++++++++++++++++++++++++++++++++++++++++
\begin{figure} % Fig. 3
\begin{center}
\resizebox{\hsize}{!}{\includegraphics[angle=90]{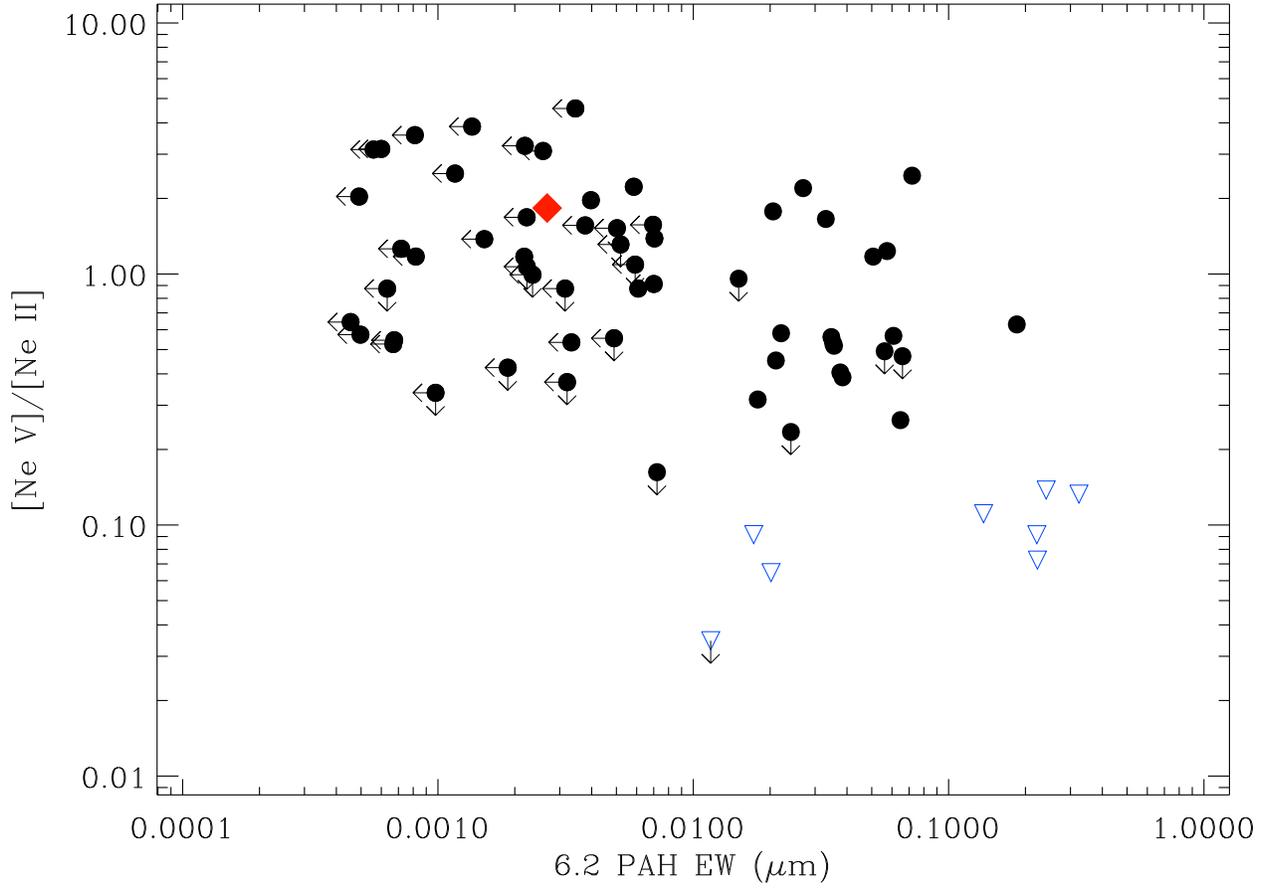}}

\caption{The EW of the 6.2$\mu m$ PAH feature versus [NeV]/[NeII]. The red diamond indicates 3C273, a well known quasar, where an AGN dominates its mid-IR spectrum (Hao et al. 2005). The blue open triangles denote objects with both high PAH emission and a low neon ratio. These are the strongest candidate starburst galaxies and were flagged as such in our sample.
\label{fig3}}

\end{center}
\end{figure}
%+++++++++++++++++++++++++++++++++++++++++++++++++++++++++++++++++++++

Even after removing the most flagrant starburst contaminants, it is evident that Fig. 1 still possesses a scatter. This is largely due to the fact that star formation activity is not a binary phenomenon where it is either overwhelmingly dominant or completely dormant. The Armus et al. (2007) and Genzel et al. (1998) figures show that the diagnostics used are continuous between the two extremes and demonstrate that while there are cases where star formation or the AGN activity dominate, in many galaxies the signature of both in the mid-IR is of similar strength. Therefore, even though we removed the galaxies where we were certain starburst activity was dominating, the remaining classified as AGN still possess some scatter in our anti-correlations. This is very likely one of the reasons why our [NeV] result is not strong. 

It is evident from the figures that the two diagnostics used are continuous
between the two extremes and demonstrate that while there are cases where star formation or the AGN activity dominate, in many galaxies the signature of both in the mid-IR is of similar strength. Therefore, even though we removed the galaxies where we were certain starburst activity was dominating, the
remaining classified as AGN still possess some scatter in our anti-correlations. This is very likely one of the reasons why our [NeV] result is not strong.

To reduce the possible contribution of star formation to the observed correlations further yet, we removed the local continuum from the analysis altogether. Laurent et al. (2000) and Nardini et al. (2008) point out that the 5.5$\mu m$ continuum is dominated by emission from dust located near the AGN torus which heated to near sublimation temperatures, with minimal contribution from star formation activity or stellar photospheric emission. Since the flux of a mid-IR line is independent of local continuum levels, the ratio of the line flux to the 5.5$\mu m$ continuum versus 5.5$\mu m$ luminosity should mitigate continuum contamination altogether. 

The result of this experiment is clear. The overall anti-correlations are better than those of Fig. 1.  All four lines display $>3\sigma$ significance and [NeII] and [NeIII] actually have $>4\sigma$ significance. The correlation values are -0.58 for [SIV], -0.65 for [NeII], -0.57 for [NeV], and -0.62 for [NeIII], with null hypothesis values of $1.03\times10^{-4}$, $1.29\times10^{-6}$, $6.41\times10^{-5}$, and $3.16\times10^{-6}$ respectively. Using this method, the slope values are $-0.32\pm0.06$ for [SIV], $-0.36\pm0.05$ for [NeII], $-0.31\pm0.06$ for [NeV], and $-0.37\pm0.06$ for [NeIII]. As expected the starburst galaxies show some of the strongest [NeII] emission in our sample and are obvious outliers. The figure further suggests that the anti-correlation is present in all four lines and is driven by the central AGN. The relationship probably only weakened by star formation in the host galaxy. These results are more meaningful than the ones from Fig. 1 principally because of the inclusion of the [NeV] line which, as mentioned before, we expect most to exhibit a Baldwin effect due to its close relation to the central AGN. Therefore, its inclusion in this analysis points to the AGN as the central driver of this effect. The analysis used in Fig. 4 is shown to be a robust way to measure the decrease in line strength with increasing AGN power and may be a better diagnostic tool than the traditional analysis used in Fig. 1 for mid-IR lines.

%+++++++++++++++++++++++++++++++++++++++++++++++++++++++++++++++++++++
\begin{figure*} % Fig. 4
\begin{center}
\resizebox{\hsize}{!}{\includegraphics[angle=90]{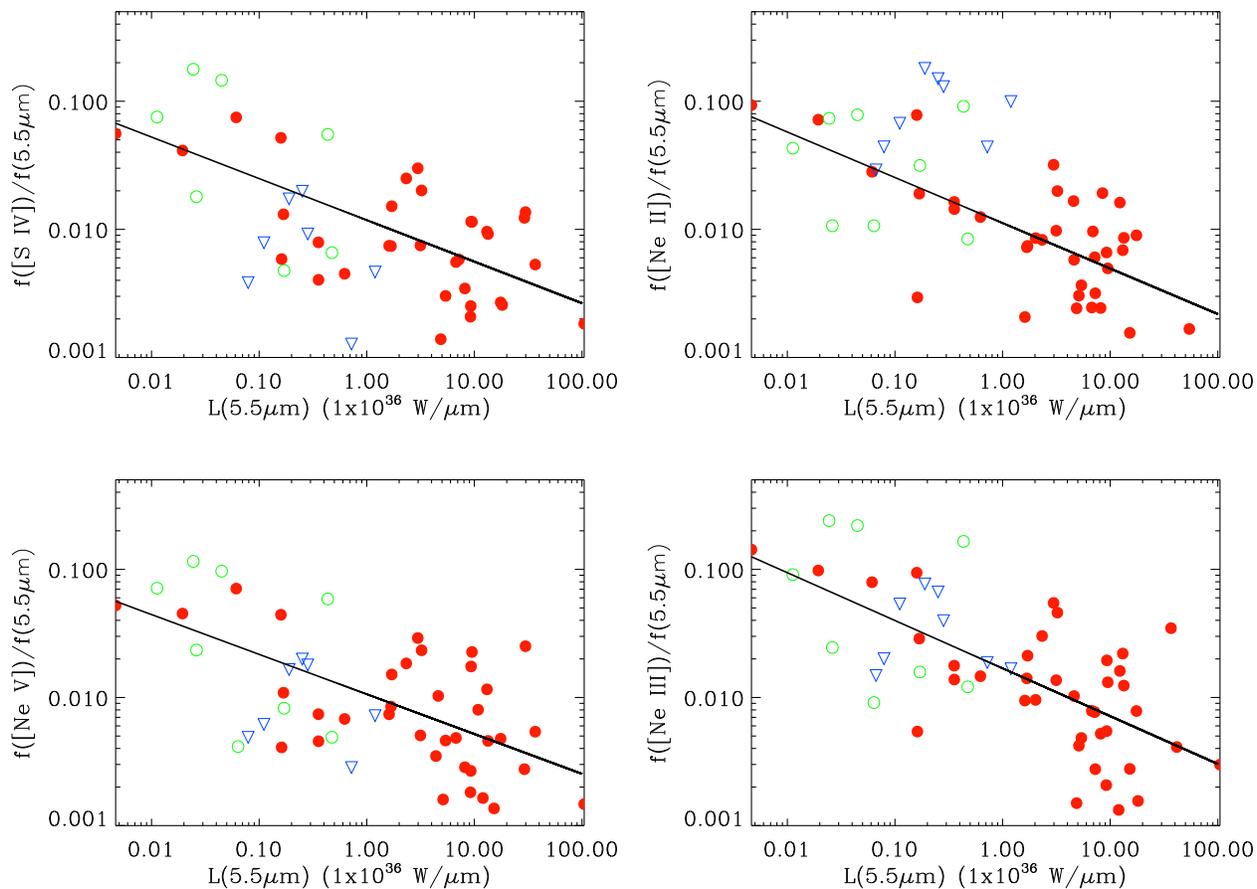}}
\caption{Plot of line flux divided by 5.5$\mu m$ flux versus 5.5$\mu m$ Luminosity for [SIV], [NeII], [NeV], and [NeIII]. The green open circles denote Seyfert 2 galaxies while the red filled circles denote Seyfert 1 galaxies. Starburst galaxies are plotted as open blue triangles. Overplotted are lines representing the best least-squares fits.
\label{fig4}}
\end{center}
\end{figure*}
%+++++++++++++++++++++++++++++++++++++++++++++++++++++++++++++++++++++

Altogether, our data support the possibility that the driver of the narrow-line Baldwin effect is a dynamic covering factor of the narrow line region. If this factor changes as a function of AGN luminosity where more luminous AGN tend to drive the NLR outward, then it would explain the decrease in relative lines strength with increasing AGN power seen in Fig. 1 and 4.

%%%%%%%%%%%%%%%%%%%%%%%%%%%%%%%%%%%%%%%%%%%%%%%%%%%%%%%%%%%%%%%%%%%%%

\acknowledgments 

We would like to thank our anonymous referee for carefully examining our manuscript and providing suggestions which greatly improved it as well as Fred Hamann for his insightful comments on the paper and for taking the time to revise the work. Dan Weedman was also very helpful in putting together this paper. VC would like to acknowledge partial support from the EU ToK grant 39965. This work is based on observations made with the Spitzer Space Telescope, which is operated by the Jet Propulsion Laboratory, California Institute of Technology, under NASA contract 1407. Support for this work by the IRS GTO team at Cornell University was provided by NASA through contract 1257184 issued by JPL/Caltech.

%%%%%%%%%%%%%%%%%%%%%%%%%%%%%%%%%%%%%%%%%%%%%%%%%%%%%%%%%%%%%%%%%%%%%


\newpage
\begin{thebibliography}{}

\bibitem[]{a1} Armus, L. et al. 2007 \apj, 656, 148
\bibitem[]{b1} Baldwin, J.A. 1977 \apj, 214, 679
\bibitem[]{b2} Baskin, A., Laor, A. 2004 MNRAS, 350, 31
\bibitem[]{b3} Boroson, T.A., Green, R.F. 1992 ApJS, 80, 109
\bibitem[]{2008ASPC..381....3C} Charmandaris, V.\ 2008, Infrared Diagnostics of Galaxy Evolution, 381, 3 
\bibitem[]{c1} Croom, S.M., et al. 2002 MNRAS, 337, 275
\bibitem[]{c2} Czerny, B., Nikolajuk, M., Piasecki, M., Kuraszkiewicz, J. 2001, MNRAS, 325, 865
\bibitem[]{d1} Dietrich M., Hamann F., Shields, J.C., et al. 2002 \apj, 581, 912\bibitem[Genzel et al.(1998)]{1998ApJ...498..579G} Genzel, R., et al.\ 
1998, \apj, 498, 579
\bibitem[]{g1} Green, P.J., Forster K., Kuraszkiewicz J. 2001 \apj, 556, 727
\bibitem[]{g2} Gorjian, V., Cleary, K., Werner, W. M., Lawrence,  C. R. 2007 \apj, 655, L73
\bibitem[]{h1} Hao, L. et al. 2005 \apj, 625, L75
\bibitem[]{h2} H\"{o}nig, S.~F., Smette, A., Beckert, T., Horst, H., Duschl, W., Gandhi, P., Kishimoto, M., \& Weigelt, G.\ 2008, A\&A 485, L21
\bibitem[Houck et al.(2004)]{2004ApJS..154...18H} Houck, J.~R., et al.\ 2004, \apjs, 154, 18 
\bibitem[]{i1} Iwasawa, K., Taniguchi, Y., 1993 \apj, 413, 15
\bibitem[]{k1a} Keremedjiev, M., \& Hao, L.\ 2006, Bulletin of the American Astronomical Society, 38, 1096 
\bibitem[]{k1} Kinney, A.L., Rivolo, A.R., Koratkar, A.P. 1990 \apj, 357, 338
\bibitem[]{k2} Kuraszkiewicz, J.K., et al. 2002 ApJS, 143, 257
\bibitem[]{l1} Laurent, O. et al. 2000 A\&A, 359, 887
\bibitem[]{l2} Lutz, D. et al. 1998 \apj, 505, L103
\bibitem[]{m1} McHardy,I.M., Gunn,K.F., Uttley,P., Goad,M.R. 2005 MNRAS, 359, 1469
\bibitem[]{2008MNRAS.385L.130N} Nardini, E., Risaliti, G., Salvati, M., Sani, E., Imanishi, M., Marconi, A.,  \& Maiolino, R.\ 2008, \mnras, 385, L130 
\bibitem[]{n1} Nandra, K., George, I.M., Mushotzky, R.F., Turner, T.J., Yaqoob, T. 1997 \apj, 488, 91
\bibitem[]{o1} Osmer, P.S., Shields, J.C. 1999 ASPC, 162, 235O
\bibitem[]{p2} Page, K.L., Turner, M.J.L., Done, C., O'Brien, P.T., Reeves, J.N., Sembay, S., Stuhlinger, M. 2004, MNRAS, 349, 57
\bibitem[]{p3} Panessa, F et al. 2006 A\&A, 455, 173
\bibitem[]{r1} Riess, A.G., et al. 2005 \apj, 627, 579
\bibitem[]{s1} Shang, H., et al. 2003 \apj, 586, 52
\bibitem[]{s2} Spoon, H. W. W. et al. 2007 \apj, 654, L49
\bibitem[]{v1} Vestergaard, M., Peterson, B.M. 2006 \apj, 641, 689
\bibitem[]{w2} Warner, C., Hamann, F., Dietrich, M. 2003 \apj, 596, 72
\bibitem[]{w3} Warner, C., Hamann, F., Dietrich, M. 2008 \emph{in prep}
\bibitem[]{w1} Wilkes, B.J., Kuraszkiewicz, J., Green, P.J., Mathur, S., McDowell, J.C. 1999, \apj, 513, 76
\bibitem[]{w2} Woo, J.H., Urry, C.M. 2002 \apj, 579, 530
\bibitem[Wu et al.(1983)]{1983ApJ...266...28W} Wu, C.-C., Boggess, A., \& Gull, T.~R.\ 1983, \apj, 266, 28 
\bibitem[]{z1} Zheng, W., Kriss, G. A., \& Davidsen, A. F.. 1995, ApJ, 440, 606

\end{thebibliography}
\end{document}